\documentclass[preprint,showpacs,preprintnumbers,amsmath,amssymb]{revtex4}
\usepackage{dcolumn}
\usepackage{bm}
\usepackage[dvipdf]{graphicx}
\DeclareGraphicsExtensions{.jpg,.pdf,.mps,.png,.eps,.ps,.EPS}
\begin{document}
\newcommand{\avg}[1]{\langle{#1}\rangle}
\newcommand{\Avg}[1]{\left\langle{#1}\right\rangle}
\def\be{\begin{equation}}
\def\ee{\end{equation}}
\def\bc{\begin{center}} 
\def\ec{\end{center}}
\def\bea{\begin{eqnarray}}
\def\eea{\end{eqnarray}}
\title{Degree distribution of complex networks from statistical mechanics principle}
\author{Ginestra Bianconi}
\affiliation{ The Abdus Salam International Center for Theoretical Physics, Strada Costiera 11, 34014 Trieste, Italy}
\begin{abstract}
 In this paper we  describe the emergence of   scale-free degree distributions from statistical mechanics principles.
We define an energy associated to a degree sequence as the logarithm of the number of indistinguishable simple networks it is 
possible to draw given the degree sequence.
Keeping fixed the total number of nodes and links, we show that the energy of  scale-free distribution is much higher than the energy  associated to the degree sequence of regular random graphs.
This results unable us to estimate the annealed average of the number of distinguishable simple graphs it is possible to draw given a scale-free distribution with structural cutoff. In particular we shaw that this number for large networks is strongly suppressed for power -law exponent $\gamma\rightarrow 2$.
\end{abstract}
\maketitle
Recently a large variety of complex systems, from the Internet to the protein interactions in the cell,  have been described in terms of the underlining complex networks \cite{RMP,Doro,Vespi}. 
In many cases the topology of these networks is  the outcome of a self-organized stochastic process since  there is no an a priori design of the connections. 
Nevertheless these structures evolve in order to perform some special task. Therefore it is crucial to ask  how much  complex networks are far from  optimal performance.
To answer this question the optimality of the networks has been defined with respect to a variety of different specific prerequisites \cite{Maritan,Flammini,Newman1,Vega,Miguel,Two_p}:  {\em i)} respect to a specific function \cite{Maritan,Flammini,Newman1},  or {\em ii)} to their dynamics \cite{Vega,Miguel}, or {\em iii)} to  some  topological robustness features \cite{Two_p}.
Nevertheless, many real networks show some universality in their topology
\cite{RMP,Doro}.  A major universal character of these networks is given by their scale-free degree distribution which is a general property of a vast class of complex systems \cite{BA}. In fact, a large variety of real networks are scale-free  with diverging  second moment of their degree distribution $\avg{k^2}$ while  others are finite-scale,  with a  finite second moment  $\avg{k^2}$. These two general network distributions have very strong impact also on the dynamics defined on these graphs as it has been widely discussed in the literature  \cite{effe}.
Until now the way to explain the appearance of the recently identified scale-free structure is two-fold: on one side there are growing network models which assume that scale-free networks dynamically emerge from a  growing process  in which new links are added following the preferential attachment rule \cite{RMP,Doro,Vespi},  on the other side there are equilibrium networks which by means of some externally imposed hidden variable present on  each node show up a  scale-free degree distribution \cite{hv,static}. 
We note here that  because of the so widely observed degree distributions of the networks, further attention must be put on their details: i.e. on the deviations from random scale-free or finite-scale distributions which makes them specific to their function as it has been  indicated for example by works on motifs and community identification \cite{Alon,Arenas}. 

In this paper we propose a method which describes the emergence of  scale-free networks  from statistical mechanics  principles.  We limit ourself to the case in  which the embedding space, if it exists,  plays a limited role in the network wiring and we allow any two nodes of the network to be linked.
We take a very simple and general assumption: we assume that the degree distribution  of the network minimize a partition function  associated to the network. 
A fundamental role in this partition function is played by the energy associated to the degree distribution. We   define this energy   as the logarithm of the  total number ${\cal N}_G$ of indistinguishable simple networks it is possible to  draw  by wiring the edges given a certain degree  distribution. Consequently, the energy is a measure of the redundancy present in the space of allowed simple networks given a degree distribution. 
This  space  is not an absolute abstract space but it is a well known space  considered  in many applications. Indeed  for example  it is sampled by the widely used randomization algorithm \cite{Sneppen}, by swapping pairs of links without changing the degree distribution.   
 The formulation of our problem assumes the aspect of some type of statistical mechanics of the networks, a direction which has been pursued also by other authors \cite{Burda,Berg,Palla,Dorogovtsev,Newmanb,Ohkubo}. However here  we don't fix  an a priori  degree distribution, or a preferential attachment dynamics like in  Ref. \cite{Burda,Dorogovtsev,Newmanb,Ohkubo} but  we take an unsupervised approach in which we  derive  the more probable distribution of a graph of $N$ nodes and $L$ links, which minimize a free energy defined in terms of a single external parameter $z$.
On the other side the free energy differs from the free energy of \cite{Palla,Berg} because the partition function is defined directly over the degree distribution. 
In this work we show that  scale-free degree distributions  with different exponent $\gamma>2$ minimize the free energy of our problem. The energy of such graphs is an extensive quantity which decrease with $\gamma$ showing that  networks  with $\gamma\rightarrow \infty$  which are regular random networks with degree distribution infinitevely peacked around their average value, have much lower energy than networks with lower $\gamma$ and diverging second moment of the connectivity $<k^2>$.
These founding  unable us to draw some conclusion on the nature of the space of simple graphs associated with given degree sequence. Making the estimation of the average number of simple graphs ${\cal N}_{SG}$ and showing that in the thermodynamic limit and when a structural cutoff is considered this number is strongly suppressed as $\gamma \rightarrow 2$.

In order to define our partition function let's consider a network with $N$ nodes and $2L$ edges with degree sequence $\{k_1,\dots,k_N\}$ associated to the degree distribution ${N_k}$.
We define the energy  $E (\{N_k\})$ associated to the degree distribution of the network as the logarithm of the  number ${\cal N}_{G}$ of indistinguishable simple networks it is possible to draw given the degree sequence, i.e. 
\be
E (\{N_k\})=\log({\cal N}_G).  
\ee
where a simple network is a graph without tadpoles and double links.
The number  ${\cal N}_G$ can be expressed as 
\be
{\cal N}_G=e^{E (\{N_k\})}={\prod_k k!^{N_k}}.
\label{S_g.eq}
\ee  
In fact, for every simple graph associated with the given distribution every  permutation of the edges departing from each node generates the same set of links. These permutations are given by $\prod_k k!^{N_k}$, and consequently we derive Eq. $(\ref{S_g.eq})$.

On the other side, the number of ways ${\cal N}_{\{N_k\}}$ in which we can distribute  $2L$ edges into any degree sequence $\{k_1,\dots, k_N\}$ of distribution $\{N_k\}$ is given by
\be
{\cal N}_{\{N_k\}}=\frac{(2L)!}{ \prod_k (k N_k)! }.
\ee
where we consider unlabeled nodes.
Thus we can define an entropy $S (\{N_k\})$ of each distribution as 
\be
e^{S (\{N_k\})}= {\cal N}_{\{N_k\}}=\frac{(2L)!}{ \prod_k (k N_k)! }.
\ee
Once we have defined the energy $E (\{N_k\})$ and the entropy $S (\{N_k\})$ associated to  a network degree distribution, we  can define  a partition function of the network.
 
Proceeding as in standard statistical mechanics, we define  a normalized partition function $Z$ of the network as the sum over all microstates of the problem, i.e. degree distributions, with given energy $E (\{N_k\})$ and entropy $S (\{N_k\})$
\be
Z=\frac{1}{(2L)!}\sum_{ \{N_k \} } {}^{'}  e^{-\frac{1}{z} E (\{N_k\})+S(\{ N_k\})}.
\label{Z.eq}
\ee
In other words  we would like to know which are the more likely distributions  $\{N_k\}$ which minimize the free energy of the network $F(\{N_k\})= E (\{N_k\}) -z S(\{ N_k\})$.
The role of the parameter  $z$ is to measure  a tradeoff between the 'energetic' and the 'entropic' term in the definition of the free energy, as well as the temperature  $T$ in classical statistical mechanics.  
In equation $(\ref{Z.eq})$ the sum $\sum'$ over the $\{N_k\}$ distributions is extended only to  $\{N_k\}$ for which  the total number of nodes $N$ and the total number of links $L$ in the network is fixed, i.e.
\bea
\sum_k N_k=N \nonumber \\
\sum_k k N_k=2 L.
\label{conditions}
\eea
To enforce these conditions we introduce in $(\ref{Z.eq})$ the delta functions in the integral form 
\be
\delta(\sum_k N_k-N)\delta(\sum_k k N_k-2 L)=\int_0^{2\pi} \frac{d\lambda}{2\pi} \int_0^{2\pi} \frac{d\nu}{2\pi}e^{i\lambda(\sum_k kN_k-2L)+i\nu(\sum_k N_k-N )}. 
\ee
Performing
\bea
Z&=&\frac{1}{(2L)!}\int \frac{d\lambda}{2\pi} \int \frac{d\nu}{2\pi} \sum_{ \{N_k \} } \exp\left[ -\frac{1}{z} E (\{N_k\})+S(\{N_k\})-i\lambda
(2L-\sum_k k N_k)-i\nu (N-\sum_k N_k)\right].\nonumber \\
&=&\int \frac{d\lambda}{2\pi} \int \frac{d\nu}{2\pi} \exp\left[-i\lambda 2L -i\nu N+  \sum_k \log G_k(\lambda,\nu)\right]=\int \frac{d\lambda}{2\pi} \int \frac{d\nu}{2\pi} \exp[Nf(\lambda, \nu)]
\label{Z2}
\eea
where
\be
G_k(\lambda, \nu)=\sum_{N_k} \frac{1}{(k N_k)!} \exp\left\{N_k\left[i\lambda k +i\nu -\frac{1}{z}\log(k!)\right] \right\}.
\ee
Assuming that the sum over all $N_k$ can be approximated by the sum over all $L_k=k N_k=1,2,\dots \infty$ we get $\log \left[G_k(\lambda, \nu)\right]=\exp\left[{i\lambda +i\nu/k -\frac{1}{kz}\log(k!)}\right]$ and
\be
f(\lambda, \nu)=-i\avg{k}\lambda-i\nu+\frac{1}{N}\sum_k e^{i\lambda+i\nu/k -\frac{1}{zk}\log(k!)},
\ee
where $<k>=2L/N$ indicates the average degree of the network.
By evaluating $(\ref{Z2})$ at the saddle point, deriving the argument of the exponential respect to $\lambda$ and $\nu$, we obtain 
\bea
\avg{k}=\frac{1}{N}\sum_k  e^{i\lambda+i\nu/k -\frac{1}{zk}\log(k!)} \nonumber \\
1=\frac{1}{N}\sum_k  \frac{1}{k} e^{i\lambda+i\nu/k -\frac{1}{zk}\log(k!)}
\label{dOmega}
\eea
and the marginal probability that  $L_k=k n=\ell$ is given by
\be
P(L_k=\ell=nk)=\frac{1}{\ell!}e^{-\ell/k\log(k!)}\frac{Z_k(L,\ell,N)}{Z(L)},
\label{marginal}
\ee
with
\be
Z_k(L,\ell,N)=\int \frac{d\lambda}{2\pi} \int \frac{d\nu}{2\pi} \exp[Nf_k(\lambda, \nu,\ell)]
\ee
and $f_k(\lambda,\nu,\ell)=i(\avg{k}-\ell/N)\lambda-i\nu(1-\ell/(kN))+\frac{1}{N}\sum_{s\neq k} \exp[{i\lambda+i\nu/s -\frac{1}{zs}\log(s!)}].$
If we develop $(\ref{marginal})$  for  $\ell \ll L $ and we use the Stirling approximation for factorials,  we get that each variable $L_k$ is a Poisson variable with mean $<L_k>$ satisfying
\be
\frac{<L_k>}{k}=<N_k>=k^{-\frac{1}{z}-1}e^{\lambda+\frac{1}{z}} e^{\nu/k}
\label{Nk}
\ee
which describe  the optimal degree distribution for our problem.
\begin{figure}
\begin{center}
\includegraphics[width=65mm, height=55mm]{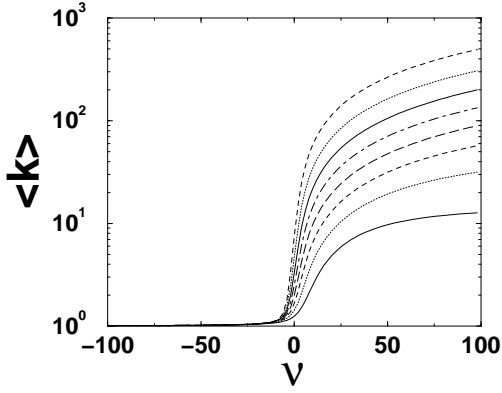}
\caption{Average degree $\avg{k}$ of the optimal distribution of as a function of $\nu$ for  networks with $z=0.2$ (solid line) $z=0.3,0.4,0.5,0.6,0.7,0.8$ and $z=0.9$ (dashed line).}
\label{kav.fig}
\end{center} 
\end{figure}
If we restrict ourself to the networks with finite average degree in the thermodynamic limit, the allowed values of $z$ are $z<1$.
From the expression $(\ref{Nk})$ of the optimal degree distribution, if $z\in(0,1)$ the optimal degree distribution is scale-free  with a power-law tail characterized by the exponent $\gamma=\frac{1}{z}+1$.  In this case  the distribution $(\ref{Nk}$) always diverges in zero, thus we necessarily must impose that there are no isolated nodes in the network. The parameter $\nu \neq 0$ modulates  the average degree of the graph constituting for $\nu>0$ an effective lower cutoff of the distribution wherever  the upper  cutoff $K$ of the degrees is  the natural cutoff of the  distribution $(\ref{Nk})$.
A different scenario arises  if $z<0$, when the optimal network $(\ref{Nk})$   has a  power-law degree distribution increasing with the degree $k$. In this case the Lyapunov functions $\nu $ and $\lambda$ cannot fix the average degree unless one introduces by hand an upper cutoff  $K$ in the degree of the nodes of the order of magnitude of the average degree $\avg{k}$. We note here that also for $z\in (0,1)$ it could be convenient to set by hand a structural cutoff $K\sim N^{1/2}$ for $z>1/2$ in order to obtain an uncorrelated network. 
In the following we will consider only the power-law case $z\in (0,1)$.

From equations $(\ref{dOmega})$   we derive for  the Lagrangian multipliers $\lambda$ and $\nu$  in   $z\in (0,1)$ :
\bea
\nu^{-\frac{1}{z}} e^{\lambda+\frac{1}{z}}\left[\Gamma(\frac{1}{z},\frac{\nu}{K})-\Gamma(\frac{1}{z},\nu) \right] &=&N\nonumber \\\nu \left[\Gamma\left(\frac{1}{z}-1,\frac{\nu}{K}\right)-\Gamma\left(\frac{1}{z}-1,{\nu}\right)\right] &= &\avg{k}\left[{\Gamma\left(\frac{1}{z},\frac{\nu}{K}\right)-\Gamma\left(\frac{1}{z},\nu\right)}\right]\nonumber \\
\nu^{-\frac{1}{z}} e^{\lambda+\frac{1}{z}}\left[\Gamma(\frac{1}{z})-\Gamma(\frac{1}{z},\frac{\nu}{K}) \right] &=&1,
\label{Pl.eq}
\eea
where $\Gamma(a,b)$ indicates the incomplete Gamma function. 
This system of equations is solvable provided $\avg{k}>1$, as can be seen from  Figure $\ref{kav.fig}$  where we plot the average degree of a network as a function of $\nu$ for different values of $z$.
\begin{figure}
\begin{center}
\includegraphics[width=65mm, height=55mm]{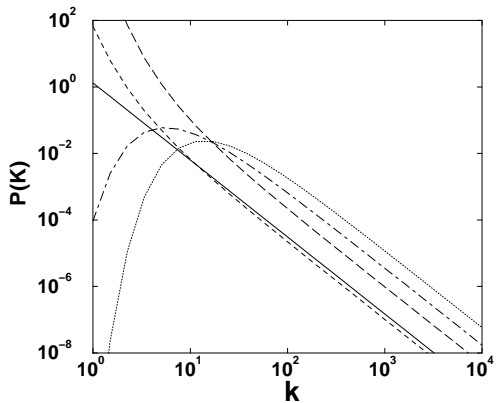}
\caption{The normalized degree distribution $P(k)$ for  $z=0.75$   as a function of the average degree of the network. Data are shown for $\avg{k}=1.3,2,4,40,100$.}
\label{stat_new.fig} 
\end{center}
\end{figure}
In fact, for  $z\in(0,1)$   there are highly connected nodes in the network and  the degree distribution for large degrees $k$ decays as a power-law with an  exponent $\gamma$ fixed by the value of $z$.
On the other side, for  small value of the degrees  $k$ the degree distribution deviates from the simple power-law and  $N_k$  depends strongly on the value of the Lagrangian multiplier $\nu$.  For low average connectivities,i.e.
\be
\avg{k}=\avg{k}_0<\frac{1}{1-z}
\ee
the solution of  Eqs. $(\ref{Pl.eq})$ involves a negative value of $\nu$. 
Accordingly, low degree nodes are more frequent than expected by a simple power-law while  for $\avg{k}>\avg{k}_0$, $\nu>0$ and the  low connected nodes are less probable than  predicted by a simple power-law.

In Figure $\ref{stat_new.fig}$ we show the optimal distributions which solve these equations  at different values of the average degree $\avg{k}$.

\begin{figure}
\begin{center}
\includegraphics[width=65mm, height=55mm]{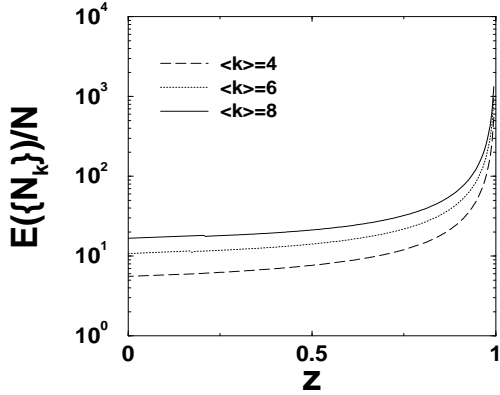}
\caption{Energy of the network as a function of the parameter $z$, for an average degree $\avg{k}=4,6,8$ in the limit $N\rightarrow \infty$. }
\label{S_g.fig} 
\end{center}
\end{figure}

Given the optimal distributions $(\ref{Nk})$  we can calculate the energy of the network as a function of $z$ at fixed average degree $\avg{k}$.
In Figure $\ref{S_g.fig}$ we present the energy of the optimal graph as a function of $z$ for different average connectivities.

The energy has a minimum in the limit  $z\rightarrow 0$ when the optimal degree distribution is  infinitely peaked around the average, and thus the graph is a random regular graph. 
On the contrary, in the limit  $z\rightarrow 1$ where  the degree distribution has a  power-law exponent $\gamma\rightarrow 2$ the energy $E (\{N_k\})$  is at the maximum.

From the derivation of our model it is evident the connection  with ``ball in the  box'' problems \cite{Ritort,Burda2,Ohkubo}. 
 In our approach the "boxes" map to   the degree of the nodes and the ``balls'' map  to the edges of the graph. This makes a crucial difference respect with the model  \cite{Ohkubo},  in which  the "boxes" map to the nodes of the graph. Although their formal similarities this difference make the two model very different in their conclusions.

We would like to indicate here that if we introduce by hand a structural cutoff $K\sim N^{1/2}$ we can assume that the networks described in this paper  are randomly wired. In this case we can evaluate the number ${\cal N}_{SG}$ of distinguishable simple graphs is it possible to construct  given the degree distribution $(\ref{Nk})$.
This number is approximated by
\be
{\cal N}_{SG}\propto\frac{(2L)!!e^{-\frac{1}{2}\left(\frac{<k^2>}{<k>}\right)^2}}{e^{E(\{N_k\})}}
\label{N_SG}
\ee
In fact the total number of wiring it  is possible to draw given $2L$ edges is given by $(2L)!!$. This number include all type of possible wiring of the edges including the ones which give rise to graphs which are not simple.
Assuming that the graph is randomly wired, i.e. that the probability that a node with $k_i$ edges connect to a node with $k_j$ edges is a Poisson variable with average  $k_i k_j/(<k>N)$ the probability $\Pi$ that the graph is simple is equal to  \cite{loops_lungo}
\be
\Pi=\prod_{i, j} \left(1+\frac{k_i k_j}{<k>N} \right)e^{-{k_i k_j}{<k>N}}\sim e^{-\frac{1}{2} \left(\frac{<k^2>}{<k>}\right)^2}.
\ee
Finally in the expression $(\ref{N_SG})$ for ${\cal N}_{SG}$ there is an additional terms which takes into account the equivalent wiring of the edges which is given by $e^{-E(\{N_k\})}$.
The term $\frac{<k^2>}{<k>}$ for scale-free graphs with cutoff $K\propto N^{1/2}$ is subleading respect to the energetic term $E{(\{N_k\})}$ which dominates for large network sizes $N$.

Consequently the total number of distinguishable simple graphs ${\cal N}_{SG}$ in an annealed approximation, is then a decreasing function of $\gamma$ suggesting  that in random scale free graphs the space of distinguishable simple random graphs is strongly suppressed as   $\gamma \rightarrow 2$. 

In conclusion, the statistical mechanics treatment of complex networks shown in this paper is able to put in a similar context, the emergence of scale-free networks and finite-scale networks. 
 Scale-free degree distribution correspond to higher energy state of the network respect to  finite-scale networks. Especially regular random graphs  have minimal energy. 
Consequently the large variety of real complex networks in many technological, biological and social systems which show a scale-free degree distribution reveal a  tendency  have a degree distribution which maximize the energy.
Furthermore we give an estimation of the annealed average size of the space of simple graphs ${\cal N}_{SG}$ for random scale-free networks showing that ${\cal N}_{SG}$ is strongly reduced as $\gamma \rightarrow 2$.

\end{document}